\documentclass[preprint,12pt]{elsarticle}

\usepackage{amssymb}

\usepackage[ruled,vlined]{algorithm2e}
\usepackage{algorithmic}
\usepackage{amsmath}

\usepackage{graphicx}

\usepackage[caption=false,font=normalsize,labelfont=sf,textfont=sf]{subfig}

\usepackage{xcolor} 
\usepackage{ulem}   
\usepackage{booktabs}

\journal{Digital Signal Processing}

\begin{document}

\begin{frontmatter}

\title{A Rate-Distortion-Classification Approach for Lossy Image Compression}

\author[inst1]{Yuefeng Zhang}

\affiliation[inst1]{organization={Beijing Institute of Computer Technology and Application},
            city={Beijing},
            country={China}}

\begin{abstract}
In lossy image compression, the objective is to achieve minimal signal distortion while compressing images to a specified bit rate. The increasing demand for visual analysis applications, particularly in classification tasks, has emphasized the significance of considering semantic distortion in compressed images. To bridge the gap between image compression and visual analysis, we propose a Rate-Distortion-Classification (RDC) model for lossy image compression, offering a unified framework to optimize the trade-off between rate, distortion, and classification accuracy. The RDC model is extensively analyzed both statistically on a multi-distribution source and experimentally on the widely used MNIST dataset. The findings reveal that the RDC model exhibits desirable properties, including monotonic non-increasing and convex functions, under certain conditions. This work provides insights into the development of human-machine friendly compression methods and Video Coding for Machine (VCM) approaches, paving the way for end-to-end image compression techniques in real-world applications.

\end{abstract}


\end{frontmatter}
\section{Introduction}

Rate-distortion theory, as the foundation of information theory, reveals the compression ratio that a source can take while maintaining credibility \cite{shannon1948mathematical}. Based on this, many research works have further explored the lower bound of the rate-distortion function, i.e., the maximum obtainable compression ratio, under a specific source distribution \cite{rufai2014lossy,Gibson2017RateDF} or real-world source distribution \cite{Yang2021TowardsES}. Nowadays, with the rise of image perceptual quality evaluation metrics (e.g.,
Learned Perceptual Image Patch Similarity (LPIPS) \cite{zhang2018unreasonable}, Fréchet Inception Distance (FID) \cite{heusel2017gans}, etc.), the rate distortion theory, which used to be optimized by Mean Squared Error (MSE) distortion or Structural Similarity Index Measure (SSIM) metrics,
has been extended to jointly optimize three optimization dimensions (i.e., rate-distortion-perception \cite{blau2019rethinking}). 

In the rate-distortion-perception \cite{blau2019rethinking} theory, the trade-off between image compression quality and bit rate is extended to the perceptual dimension. According to the rate-distortion-perception \cite{blau2019rethinking}, the improvement of the signal perception evaluation index at a certain bit rate is at the expense of information fidelity. Then, the current lossy image coding methods, which emphasize perceptual evaluation as the driving force, may cause problems in the practical application of reconstructed images. Reconstructed images cannot be accurately analyzed for some specific image applications \cite{broder1996pattern,rusyn2016lossless,mahajan2023image} (e.g., recognition of reconstructed handwritten text i or authentication of reconstructed face images). Thus, the relationship between image coding and visual analysis need to be explored.

With the widespread application of intelligent visual technologies, the concept of Video Coding for Machine (VCM) \cite{duan2020video} has been proposed, considering both human and machine analysis.
To explore the impact of signal degradation on the understanding of reconstructed signal analysis, Liu et al. \cite{liu2019classification} explored the classification-distortion-perception relationship in the field of signal reconstruction (image denoising and super-resolution), using the classification task as a representative for the image visual analysis task, and confirmed the existence of a mutual constraint between the three. However, this work mainly focused on the signal reconstruction problem without touching the important constraint of the coding problem, i.e., the rate.

Built upon the foundations of rate-distortion theory and incorporating constraints on visual analysis of reconstructed images, the motivation behind the Rate-Distortion-Classification (RDC) model is to bridge the gap between lossy image compression and machine visual analysis from both theoretical modeling and experimental analysis perspectives. By considering the classification task as a proxy for visual analysis, the RDC model simplifies the modeling process of the impact of signal degradation on the visual analysis task. This joint optimization acknowledges the trade-off between rate, perception and fidelity and aims to strike a balance between these factors.
A preliminary version of this manuscript has been published in DCC 2023 \cite{zhang2023rate} and this manuscript provides a comprehensive elaboration of the proposed model and presents additional experimental results.

The main contributions of this paper can be summarized as follows:

\begin{itemize}

\item Reviewing the existing theoretical research on the topic addressed in this paper and identifying the areas that require further investigation.

\item Modeling the RDC model to describe the connection between image compression and the effectiveness of visual analysis of reconstructed images.

\item Demonstrating the statistical characteristics of the RDC model using both a specific Bernoulli distribution source and a more general distribution source.

\item Analyzing the experimental results of the RDC model's statistical properties using the MNIST dataset.

\end{itemize}

\section{Related Work}

In the following section, three theories closely related to the content of this paper: distortion-perception \cite{blau2018perception}, rate-distortion-perception \cite{blau2019rethinking} and classification-distortion-perception \cite{liu2019classification} are sorted out.

\subsection{Distortion-Perception}
Blau et al. \cite{blau2018perception} demonstrate mathematically that the distortion-perception function relationship between distortion (measured by measures such as Peak Signal-to-Noise Ratio (PSNR), SSIM, etc.) and human eye perception in the field of image reconstruction has a convex function property.

Assuming that $X$ is the original image, $\hat{X}$ is the degraded reconstructed image, and $Y$ is the degraded signal, then the distortion-perception can be modeled optimally as follows, where the optimization objective is the perceived quality of the reconstructed signal and the constraint is the signal distortion.

\noindent \textbf{Definition 1.} Distortion-perception in signal restoration task is defined as:
\begin{equation}
P(D)=\min _{p_{\hat{X} \mid Y}} d\left(p_{X}, p_{\hat{X}}\right) \quad \text { s.t. } \quad \mathbb{E}[\Delta(X, \hat{X})] \leq D
\label{eqa:p-d}
\end{equation}
where $\Delta(\cdot,\cdot)$ is the distortion metric, $d(\cdot,\cdot)$ evaluates the different between distortions and $d(p_{X},p_{\hat{X}})$ is the metric for signal perceptual quality.

\subsection{Rate-Distortion-Perception}
Blau et al. \cite{blau2019rethinking} combined the above distortion-perception function with information theory to model the joint rate-distortion-perception optimization by considering the perception dimension into the rate-distortion function, where the optimization objective function is the code rate expressed in terms of the mutual information $I(\cdot,\cdot)$, and the constraints are the image distortion and the image perceptual quality.

\noindent \textbf{Definition 2.} Rate-distortion-perception is defined as:
\begin{equation}
\begin{aligned}
R(D, P)=& \min _{p_{\hat{X} \mid X}} I(X, \hat{X}) \\
& \text { s.t. } \mathbb{E}[\Delta(X, \hat{X})] \leq D, d\left(p_{X}, p_{\hat{X}}\right) \leq P
\end{aligned}
\label{eqa:r-p-d}
\end{equation}
where $P$ is the constraint condition for perception.

\subsection{Classification-Distortion-Perception}

Considering the application of degraded reconstructed signals in analytical recognition tasks, Liu et al. \cite{liu2019classification} extended the distortion-perception function to the machine vision analysis dimension, specifically using the classification error rate in classification tasks as an evaluation metric.

To facilitate the derivation of the illustration, Liu et al. \cite{liu2019classification} selected the binary classification problem as a representative metric to analyze the performance on degraded reconstructed images. It is assumed that the original signal belongs to one of two classes: $\omega_1$ or $\omega_2$. Define the prior probability and probability mass function as $P_1, P_2 = 1 - P_1$ and $p_{X1}(x), p_{X2}(x)$, respectively. Then original signal $X$ satisfies $p_X(x) = P_1 p_{X1}(x) + P_2 p_{X2}(x)$. Degradation signal $Y$ has $p_Y(y) = P_1 p_{Y1}(y) + P_2 p_{Y2}(y)$ and reconstructed signal $\hat{X}$ has $p_{\hat{X}}(\hat{x}) = P_1 p_{\hat{X}_1}(\hat{x}) + P_2 p_{\hat{X}_2}(\hat{x})$ that
\begin{equation}
\begin{aligned}
&p_{Y i}(y)=\sum_{x \in \mathcal{X}} p(y \mid x) p_{X i}(x), i=1,2 \\
&p_{\hat{X} i}(\hat{x})=\sum_{y \in \mathcal{Y}} p(\hat{x} \mid y) p_{Y i}(y)=\sum_{y} \sum_{x} p(\hat{x} \mid y) p(y \mid x) p_{X i}(x), i=1,2.
\end{aligned}
\end{equation}

Define the binary classifier as:

\begin{equation}
c(t)=c(t \mid \mathcal{R})= 
\begin{cases}
    \omega_{1}, & \text { if } t \in \mathcal{R} \\ 
    \omega_{2}, & \text { otherwise }
\end{cases}
\label{eqa:binary-cls}
\end{equation}
where $\mathcal{R}$ is the set of values of $t$ corresponding to the category $\omega_{1}$. The classification error rate obtained from the input binary classifier can be expressed as:
\begin{equation}
\begin{aligned}
\text { Classification Error Rate } &:=\varepsilon(\hat{X} \mid c)=\varepsilon(\hat{X} \mid \mathcal{R}) \\
&=P_{2} \sum_{\hat{x} \in \mathcal{R}} p_{\hat{X} 2}(\hat{x})+P_{1} \sum_{\hat{x} \notin \mathcal{R}} p_{\hat{X} 1}(\hat{x})
\end{aligned}
\label{eqa:class-error}
\end{equation}

Based on the above two-class classifier and the definition of the classification error rate, Liu et al. \cite{liu2019classification} model the following joint classification-distortion-perception optimization in the signal degradation task, where the optimization objective is the classification error rate on the reconstructed image signal and the constraints are the image distortion and image perceptual quality.

\noindent \textbf{Definition 3.} Classification-distortion-perception is defined as:
\begin{equation}
\begin{aligned}
C(D, P)=& \min _{P_{\hat{X} \mid Y}} \varepsilon\left(\hat{X} \mid c_{0}\right), \\
& \text { s.t. } 
\mathbb{E}[\Delta(X, \hat{X})] \leq D, d\left(p_{X}, p_{\hat{X}}\right) \leq P
\end{aligned}
\label{eqa:c-d-p}
\end{equation}
where $c_{0}=c\left(\cdot \mid \mathcal{R}_{0}\right)$ is the predefined binary classifier on the reconstructed image.

\section{RDC and Multi-distribution Source Analysis}
\label{sec:rdc}
Referring to the above-mentioned work, we re-start from the information-theoretic rate-distortion theory and extend it to the perceptual understanding dimension, i.e., to be measured by the classification error rate. To achieve a joint optimal modeling of image compression and visual analysis of reconstructed images, we first define the RDC optimization modeling. Then, we analyse the statistical properties possessed by RDC model according to the original signals belonging to Bernoulli-distributed sources and general-distributed sources, respectively.

\subsection{RDC Joint Optimization Modeling}

For the joint optimization modeling of rate-distortion-classification, the optimization objective function is the code rate expressed by the mutual information $I(\cdot, \cdot)$ with the constraints MSE loss $\mathbb{E}[\Delta(\cdot, \cdot)]$ and the classification error rate $\varepsilon$, where $\varepsilon$ is defined by the reference Eq.~(\ref{eqa:class-error}).

\noindent \textbf{Definition 4.} In lossy image compression, RDC is defined as:
\begin{equation}
\begin{aligned}
I(X,\hat{X})=& \min _{P_{\hat{X} \mid X}} I(X, \hat{X}), \\
& \text { s.t. } 
\mathbb{E}[\Delta(X, \hat{X})] \leq D, \varepsilon(\hat{X} \mid C_0) \leq E
\end{aligned}
\label{eqa:r-d-e}
\end{equation}
where $C_{0}=c\left(\cdot \mid \mathcal{R}_{0}\right)$ is the predefined binary classifier.



\subsection{Bernoulli Source}
Assuming that the binary signal $X \sim Bern(p)$ is encoded, the reconstructed signal $\hat{X}$ is also a binary signal. The distortion measure $\delta(\cdot, \cdot)$ is assumed to be the Hamming Distance, and $p \leq \frac{1}{2}$. When the reconstruction signal classification error rate is unconstrained (i.e. $E = \infty$), then Eq. (\ref{eqa:r-d-e}) degenerates to a rate-distortion function of the binary signal source \cite{cover1999elements}, which can be expressed as:
\begin{equation}
R(D, \infty)= 
\begin{cases}
    H_{b}(p)-H_{b}(D) & D \in[0, p) \\ 
    0 & D \in[p, \infty)
\end{cases}
\label{eqa:p=infty}
\end{equation}
where $H_b(\alpha)$ is the information entropy for Bernoulli distribution with the probability of $\alpha$\footnote{$H_b(\alpha) = -\alpha \log(\alpha) - (1-\alpha)\log(1-\alpha).$}.

Then RDC optimization problem can be solved as:
\begin{equation}
R(D, E)= 
\begin{cases}
H_{b}(p)-H_{b}(D) & D \leq D_{1} \\ 
I_1 & D_{1}<D \leq D_{2} \\
0 & D_{2}<D
\end{cases}
\label{eqa:bern-answer}
\end{equation}
where $I_1$ is the notion for corresponding symbolic solution. The detailed calculation of Eq. (\ref{eqa:bern-answer}) and numerical representation of $I_1$ can be calculated by scientific computing software.
When the original binary signal obeys Bernoulli distribution, the analytical solution of RDC in Eq. (\ref{eqa:r-d-e}) can be obtained as Eq. (\ref{eqa:bern-answer}). 

\subsection{General Source}

For subsequent analysis on general distribution sources, we first set the following assumptions on the error function $\varepsilon$ for the binary classifier.

\noindent \textbf{Assumption 1.} If $\varepsilon(p \mid c_0)$ is convex on argument $p$, for any $\lambda \in [0,1]$ exists $p_1, p_2$ which satisfy:
\begin{equation}
\varepsilon(\lambda p_1 + (1-\lambda) p_2 \mid c_0) \leq \lambda \varepsilon(p_1 \mid c_0) + (1-\lambda)\varepsilon(p_2).
\end{equation}

\noindent \textbf{Theorem 1.} Considering RDC function $R(D,E)$ in Eq. (\ref{eqa:r-d-e}):
\begin{enumerate}
\item is monotonically non-increasing in $D$ and $E$;
\item is convex if \textbf{Assumption 1} holds.
\end{enumerate}

\noindent \textit{Proof.} In the following, two properties of the RDC: monotonicity and convexity, are proved one by one.

\noindent  \textbf{monotonicity} $R(D,E)$ is the minimum mutual information on a finite set and this finite set increases as the variables $D,E$ increase, i.e., it is proved that $R(D,E)$ is monotonically non-increasing on the variables $D$ and $E$.

\noindent \textbf{Convexity} Assume that \textbf{Assumption 1} holds. Proving that $R(D,E)$ has convex functionality is equivalent to proving that for any $\lambda \in [0,1]$ satisfies:
\begin{equation}
\lambda R(D_1,E_1) + (1-\lambda) R(D_2,E_2) \geq R(\lambda D_1 + (1-\lambda) D_2, \lambda E_1 + (1-\lambda) E_2),
\label{eqa:r-d-e-convex}
\end{equation}
The left side of the inequality can be written as:
\begin{equation}
\lambda I(X, \hat{X}_1) + (1-\lambda) I(X, \hat{X}_2),
\label{eqa:lambda}
\end{equation}
where $\hat{X}_1$ and $\hat{X}_2$ are defined as:

\begin{gather}
p_{\hat{X}_{1} \mid X}=\arg \min _{p_{\hat{X} \mid X}} I(X, \hat{X}) \quad \text { s.t. } \quad \mathbb{E}[\Delta(X, \hat{X})] \leq D_{1}, \varepsilon(\hat{X} \mid c_0) \leq E_1, \\
p_{\hat{X}_{2} \mid X}=\arg \min _{p_{\hat{X} \mid X}} I(X, \hat{X}) \quad \text { s.t. } \quad \mathbb{E}[\Delta(X, \hat{X})] \leq D_{2}, \varepsilon(\hat{X} \mid c_0) \leq E_2.
\end{gather}
Since $I(X, \hat{X})$ is convex when $p_X$ is fixed, we have:
\begin{equation}
\lambda I(X, \hat{X}_1) + (1-\lambda) I(X, \hat{X}_2) \geq I(X, \hat{X}_{\lambda}),
\label{eqa:lamba-geq}
\end{equation}
where $\hat{X}_{\lambda}$ follows the distribution as:
\begin{equation}
P_{\hat{X}_{\lambda} \mid X} = \lambda P_{\hat{X}_{1} \mid X} + (1- \lambda) P_{\hat{X}_{2} \mid X}.
\label{eqa:x-lambda}
\end{equation}
Denote $D_{\lambda}=\mathbb{E}\left[\Delta\left(X, \hat{X}_{\lambda}\right)\right]$ and $P_{\lambda}=d(p_{X}, p_{\hat{X}_{\lambda}})$, then we have:
\begin{equation}
I\left(X, \hat{X}_{\lambda}\right) \geq \min _{p_{\hat{X} \mid X}}\left\{I(X, \hat{X}): \mathbb{E}[\Delta(X, \hat{X})] \leq D_{\lambda}, \varepsilon(\hat{X}_{\lambda} \mid c_0) \leq E_{\lambda} \right\} = R\left(D_{\lambda}, E_{\lambda}\right),
\end{equation}
where $\hat{X}_{\lambda}$ acts on a finite set. By \textbf{Assumption 1} it is known that $\varepsilon(p \mid c_0)$ is convex, then:
\begin{equation}
\begin{aligned}
E_{\lambda} &=
\varepsilon(\hat{X}_{\lambda} \mid c_0) \\
& \leq \lambda \varepsilon(\hat{X}_1 \mid c_0) + (1-\lambda) \varepsilon (\hat{X}_2 \mid c_0) \\
& \leq \lambda E_1 + (1-\lambda)E_2
\end{aligned}
\label{eqa:e-convex}
\end{equation}
Similarly we have:
\begin{equation}
\begin{aligned}
D_{\lambda} &=\mathbb{E}\left[\Delta\left(X, \hat{X}_{\lambda}\right)\right] \\
& \stackrel{(\mathrm{a})}{=} \mathbb{E}\left[\mathbb{E}\left[\Delta\left(X, \hat{X}_{\lambda}\right) \mid X\right]\right] \\
& \stackrel{(\mathrm{b})}{=} \mathbb{E}\left[\lambda \mathbb{E}\left[\Delta\left(X, \hat{X}_{1}\right) \mid X\right]+(1-\lambda) \mathbb{E}\left[\Delta\left(X, \hat{X}_{2}\right) \mid X\right]\right] \\
& \stackrel{(\mathrm{c})}{=} \lambda \mathbb{E}\left[\Delta\left(X, \hat{X}_{1}\right)\right]+(1-\lambda) \mathbb{E}\left[\Delta\left(X, \hat{X}_{2}\right)\right] \\
& \leq \lambda D_{1}+(1-\lambda) D_{2}
\end{aligned}
\label{eqa:d-convex}
\end{equation}
where (a) and (c) are the Law of Total Expectation and (b) is derived from Eq. (\ref{eqa:x-lambda}) \cite{blau2019rethinking}. Since $R(D, E)$ is non-increasing on $D$ and $E$, it follows from equation (\ref{eqa:e-convex}) and equation (\ref{eqa:d-convex}) that:
\begin{equation}
R(D_{\lambda}, E_{\lambda}) \geq R(\lambda D_1 + (1-\lambda)D_2, \lambda E_1 + (1-\lambda)E_2).
\label{eqa:r-d-e-lambda}
\end{equation}
From Eq. (\ref{eqa:lambda}), Eq. (\ref{eqa:lamba-geq}), Eq. (\ref{eqa:x-lambda}), and Eq. (\ref{eqa:r-d-e-lambda}), we can deduce Eq.~(\ref{eqa:r-d-e-convex}), which proves that $R(D, E)$ conforms to the convex property.

\section{Experiments}

\subsection{Experiment Settings}
\subsubsection{Model Settings}

The auto-encoder structure \cite{hinton2006reducing} is used as the lossy image coding network in the experiments which architecture details are shown in Table~\ref{table:autoencoder_architecture}.
We assume that the encoder is $E$, decoder is $D$, and input image is $X$. We have interval coding, i.e., the output of encoder $E(X)$ will be mapped to $L$ intervals so the compression bit-rate will be $d \log{L}$, where $d$ is the vector dimension of the encoder's output. To ensure that gradient calculation is effective in training phase, we have dithered quantization \cite{Gray1991DitheredQ,Gray1993DitheredQ} to approximate the quantified values. In all, the encoder side will calculate:
\begin{gather}
Z = \mathcal{Q}(E(X) + u) \\
u \sim U [-\frac{1}{L-1}, \frac{1}{L-1}]^d
\end{gather}
where $\mathcal{Q}$ is the scalar quantization operation which can be defined as:
$$Q(x) = \text{round}\left(\frac{x}{\Delta}\right) \cdot \Delta$$
where $\mathcal{Q}(x)$ represents the quantized value of $x$, $\Delta$ denotes the quantization step size and $\text{round}(\cdot)$ denotes the rounding function. Specifically, soft quantization \cite{tschannen2018deep} with differentiable approximation is used for gradient backpropagation.
$u$ is the random noise subjecting to distribution of $U [-\frac{1}{L-1}, \frac{1}{L-1}]^d$, $Z$ is the encoded feature. On decoder side, we reconstruct original image $X$ by receiving $Z$ and $u$.

The training loss function consists of two components: distortion loss and classification loss. The distortion loss is the MSE loss; the classification loss is the negative log likelihood loss corresponding to the classification of the compressed reconstructed image by the neural network pre-trained on the MNIST dataset, set to $e$. Then the loss function corresponding to the training is as:
\begin{gather}
\mathcal{L} = \lambda d(X, \hat{X}) + e(\hat{X}) \\
d(\cdot, \cdot) = \mathbb{E}[||\cdot, \cdot||^2]
\end{gather}
where $d(\cdot,\cdot)$ denotes the MSE loss and $\lambda$ is the hyperparameter to balance the two parts of the total loss. In experiments, we change the model compression rate by adjusting the number of quantization intervals $L$.

\begin{table*}[t]

\caption{
Network architecture of the autoencoder. $L(x,y)$ represents a linear layer with $x$ dimension input and $y$ dimension output. $BN(x)$ is defined as one-dimensional batch normalization with $x$ dimension input. T-Conv is the vanilla transposed-convolution, e.g.$\text{T-Conv}(3\times3, 256, 2)$ is the  transposed-convolution with kernel size $3\times3$, convolution dimension $256$ and stride $2$. $N$, $d_{input}$ and $d_{latent}$ are dimension sizes. LReLU() represents 
Leaky ReLU activation function.
}

\begin{center}
\begin{tabular}{@{}ll@{}}
\toprule
\textbf{Encoder} & \textbf{Decoder} \\ \midrule
L($d_{input}$, 4N), BN(4N), LReLU() & L($d_{latent}$, 128), BN(128), LReLU() \\
L(4N, 2N), BN(2N), LReLU() & L(128, 512), BN(512), LReLU() \\
L(2N, N), BN(N), LReLU() & T-Conv(5x5, 64, 2), BN(64), LReLU() \\
L(N, N), BN(N), LReLU() & T-Conv(5x5, 128, 2), BN(128), LReLU() \\
L(N, $d_{latent}$), BN($d_{latent}$), Tanh() & T-Conv(4x4, 1, 1), Sigmoid() \\ \bottomrule
\end{tabular}
\end{center}

\label{table:autoencoder_architecture}
\end{table*}

\begin{table*}[t]

\caption{
Network architecture for the pretrained classifier network. max\_pool represents max pooling operation with the stride of 2. ReLU strands for Rectified Linear Unit. Flatten is the operation to convert the feature into one dimension. Dropout probability is set as $0.5$.
}

\begin{center}
\begin{tabular}{@{}ll@{}}
\toprule
\textbf{Layer} & \textbf{\#Parameters} \\ \midrule
Conv(5x5, 1, 10), max\_pool, ReLU & 260 \\
Conv(5x5, 10, 20), max\_pool, ReLU & 5,020 \\
Flatten & - \\
L(320, 50), ReLU & 16,050 \\
Dropout & - \\
L(50, 10), log\_softmax & 510 \\ \bottomrule
\end{tabular}
\end{center}

\label{table:classifier_architecture}
\end{table*}

\subsubsection{Dataset}
The experiments in this paper are validated on the MNIST dataset \cite{LeCun2005TheMD}, which is a handwritten digital image dataset with 60,000 images in the training set and 10,000 images in the test set. The images are in black-and-white format with a fixed size of 28 × 28 pixels, and the numbers are preprocessed to the center of the images; the labels correspond to Arabic numerals 0 to 9.

\subsubsection{Training Settings}
In training, we adopt hyperparameter $\lambda$ belonging to the set \{0, 0.0033, 0.005, 0.0066, 0.008, 0.01, 0.011, 0.013, 0.015\}. Quantization intervals $L$ is from set \{2,3,4,5\}. Batch size is set as $64$ and the model is trained for $30$ epochs. The learning rate is fixed at $0.01$.
Network channel numbers N, $d_{input}$, $d_{latent}$ are 128, 784 and 3, respectively.

For the pretrained classification network, we train the convolution network for $20$ epochs with learning rate $0.01$, which network is shown in Table \ref{table:classifier_architecture}. 
Batch size is set as $64$. The pretrained network achieves the classification accuracy of $98.7\%$ on test split.

\subsection{Result Analysis}

We first analyze the training loss of the RDC model, which indicates the convergence status of the model training. Next, we examine the characteristics of the RDC model in the experiments and compare the experimental results with the mathematical proof conclusions mentioned in Section \ref{sec:rdc}.

\begin{figure}[t]
  \centering
  \includegraphics[width=1.0\textwidth]{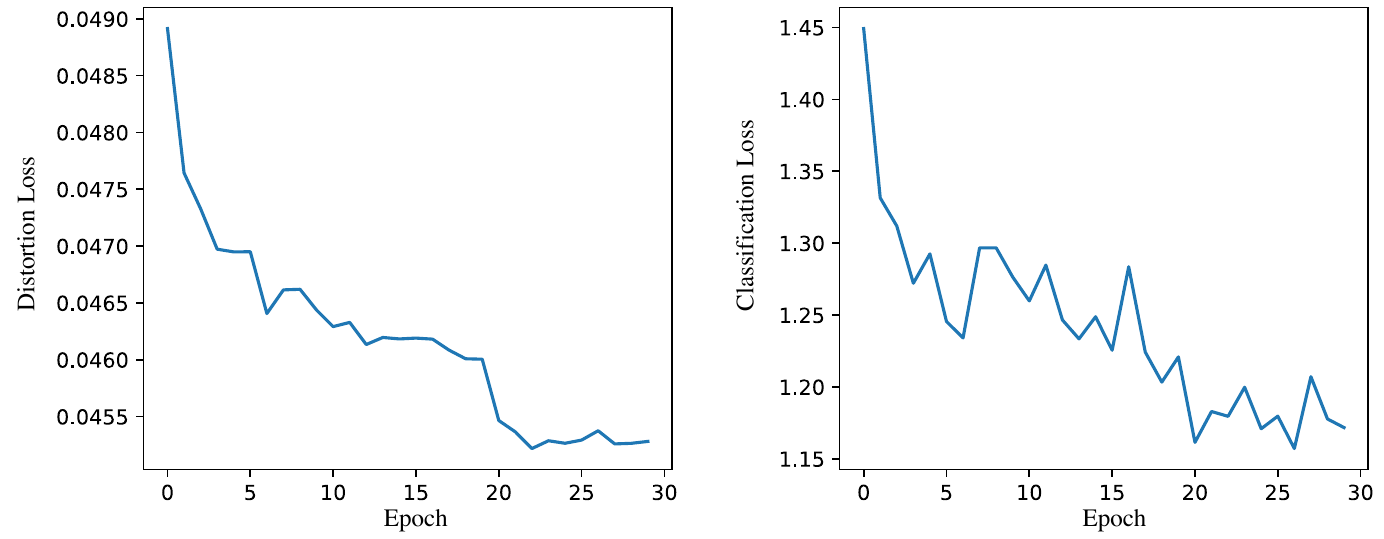}
  \caption{Training loss diagram with $L=3,\lambda=0.015$. \textbf{Left:} Distortion loss. \textbf{Right:} Classification loss.}
  \label{fig:d-c-losses}
\end{figure}

\begin{figure}[t]
  \centering
  \includegraphics[width=0.88\textwidth]{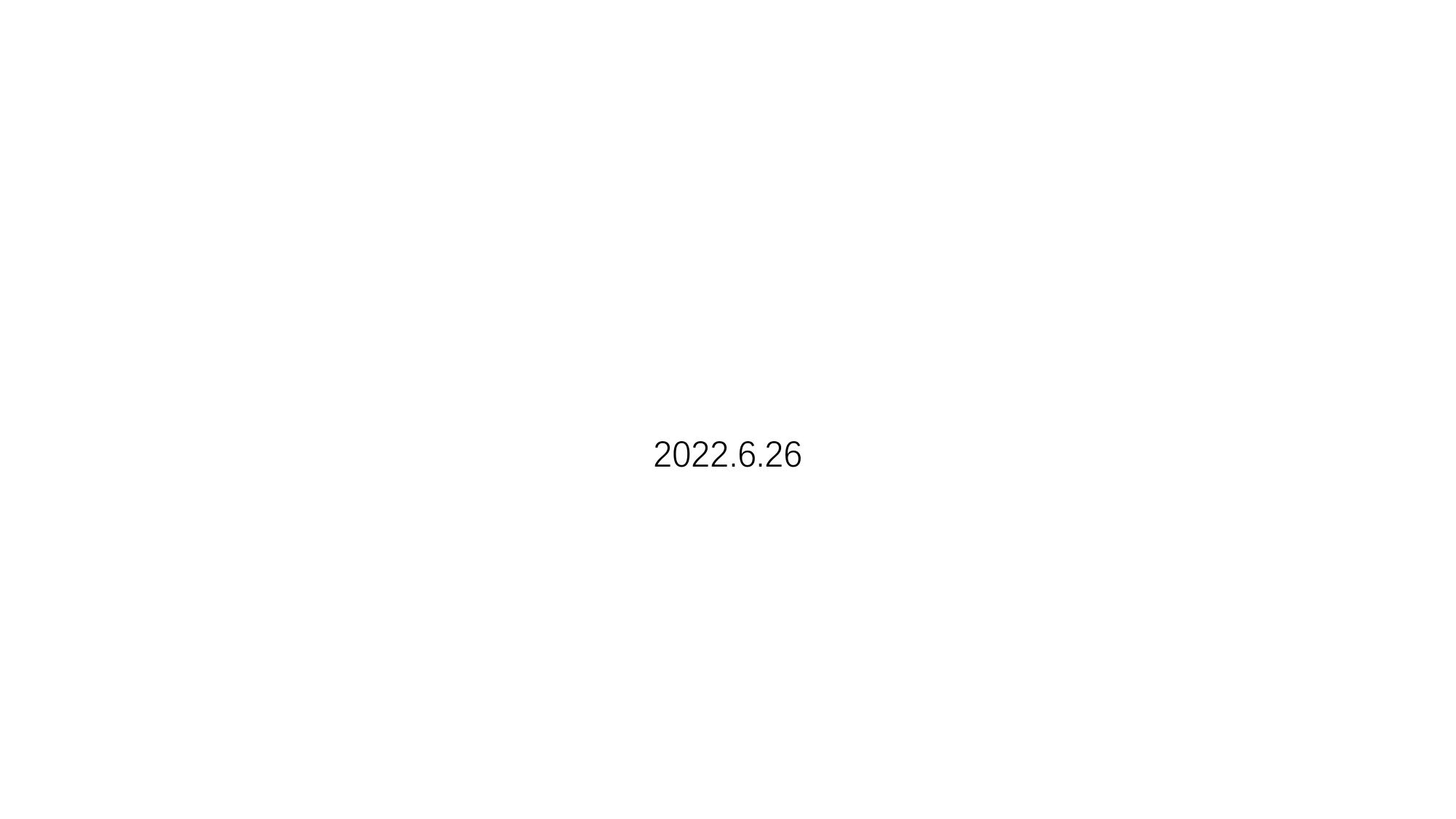}
  \caption{Reconstructed images from the compression model with different $\lambda$.}
  \label{fig:recon}
\end{figure}

\subsubsection{Training Loss}

We use the model with $L=3$ and $\lambda=0.015$ as an example, and the distortion loss and classification loss during the training phase are shown in Figure \ref{fig:d-c-losses}. From the figure, we can observe that the training process effectively converges at epoch 30. Furthermore, the trend of the loss depicted in Figure \ref{fig:d-c-losses} indicates that the distortion loss and classification loss both decrease simultaneously as the training iterations increase, for a specific quantization level of the same model. This result suggests that optimizing the reconstruction of pixel-level distortion aligns with improving visual analysis performance under a fixed target compression rate. This finding supports the mathematical conclusions presented in Section \ref{sec:rdc}.

\begin{figure}[t]%
\centering
\subfloat[\centering ]{
    {\includegraphics[width=0.49\linewidth]{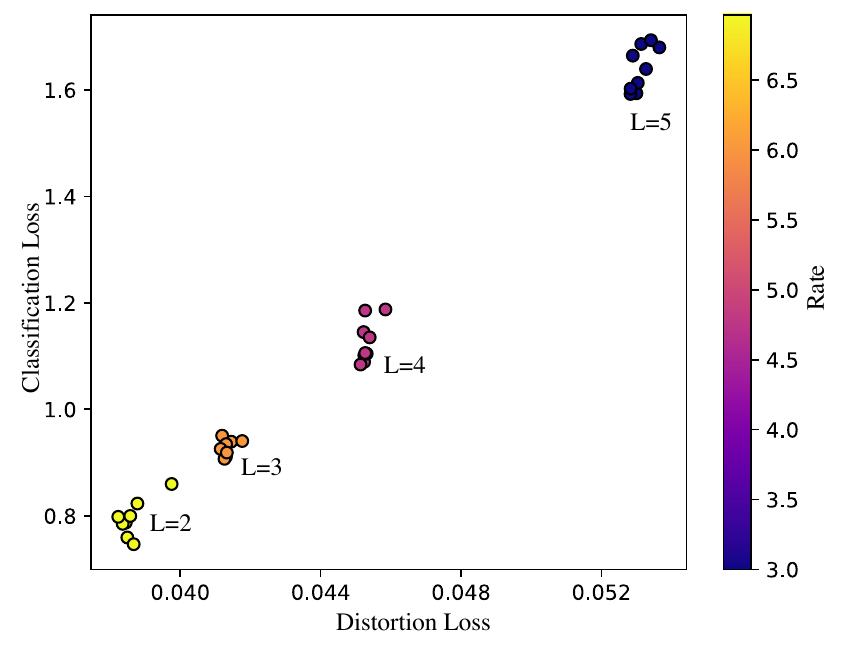} }
}
\subfloat[\centering ]{
    {\includegraphics[width=0.48\linewidth]{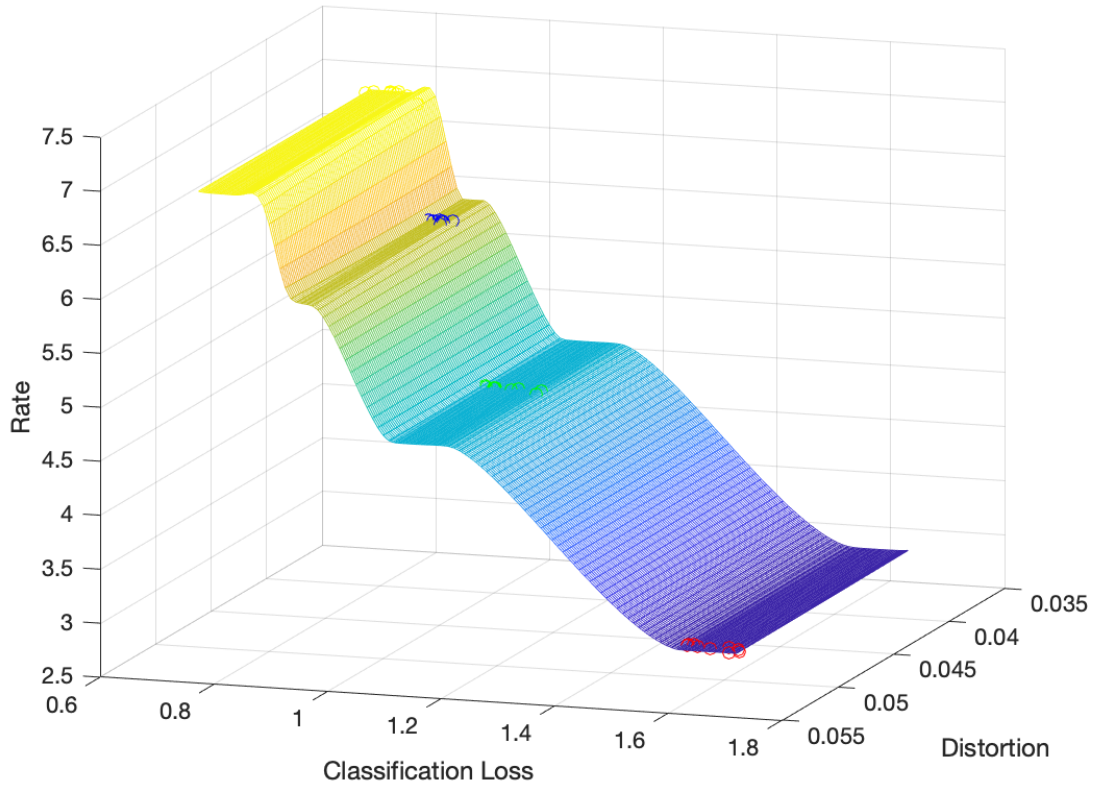} }
}

\caption{(a) RDC two-dimensional diagram. (b) RDC three-dimensional diagram. }

\label{fig:rdc-tradeoff}
\end{figure}

\subsubsection{RDC Analysis}

The impact of varying the quantization interval $L$ and the loss balance hyperparameter $\lambda$ on the reconstructed image sharpness is illustrated in Fig.~\ref{fig:recon}. As $\lambda$ increases, the reconstructed image becomes progressively sharper.
To evaluate different models, we conducted tests on the training set and analyzed the results, which are presented in Fig.~\ref{fig:rdc-tradeoff}(a). As the rate increases, both the distortion loss and the classification loss show a simultaneous decrease. Notably, for a specific rate (corresponding to a certain number of quantization intervals $L$), the distortion loss exhibits a proportional relationship with the classification loss.  It should be noted that the same rate point converges to a different equilibrium point due to the random jitter noise, which is reflected by the different distortion and classification task losses for the same rate.

To further understand the relationship between rate, distortion, and classification, we visualize the experimental results in a three-dimensional representation depicted in Fig.~\ref{fig:rdc-tradeoff}(b). The figure demonstrates that higher rates correspond to lower classification loss and pixel-level distortion loss. These findings align with the expected monotonicity and convex function properties of RDC model on general distribution sources, and the experimental results remain consistent with the previous analysis.

\section{Conclusion}

We propose a joint RDC optimization model, and then analyze the relationship between compression and visual analysis tasks. This paper uses the classification task without loss of generality as a visual analysis task representative, where the classification task input is a compressed reconstructed image. We analytically verify the statistical properties possessed by the proposed joint RDC optimization function from the special case in statistical sense to the general case. Subsequently, the experimental results on MNIST dataset are displayed to illustrate that in lossy image coding, when the RDC conforms to the monotonicity and convex function properties, then lower rates lead to higher distortion loss and classification errors.

The analysis illustrated in this paper could provide some reference implications for lossy image coding methods oriented to visual analysis. Furthermore, the future scope of our present work includes extending the RDC model to different image modalities, exploring advanced compression techniques, and evaluating its performance in real-world applications such as medical imaging and autonomous systems.

\section*{Acknowledgements}

We thank Prof.Siwei Ma (Peking University) for valuable discussion and support. We thank Zhimeng Huang and Chuanmin Jia for assistance in experiment set-up and comments on the manuscript.

 \bibliographystyle{elsarticle-num} 
 \bibliography{references}





\end{document}